# INTELLIGENT FATIGUE DETECTION AND AUTOMATIC VEHICLE CONTROL SYSTEM

Miss. Monali Gulhane[1] and Prof. P. S. Mohod[2]

[1] PG Scholar Department of CSE GHRCOETW, RTMN University, Nagpur[Maharashtra], India

[2] Hod, Department of CSE GHRCOETW, RTMN University, Nagpur[Maharashtra], India

## *ABSTRACT*

*This paper describes method for detecting the early signs of fatigue in train drivers. As soon as the train driver is falling in symptoms of fatigue immediate message will be transfer to the control room indicating the status of the drivers. In addition of the advance technology of heart rate sensors is also added in the system for correct detection of status of driver if in either case driver is falling to fatigue due to any sever medical problems .The fatigue is detected in the system by the image processing method of comparing the image(frames) in the video and by using the human features we are able to estimate the indirect way of detecting fatigue. The technique also focuses on modes of person when driving the train i.e. awake, drowsy state or sleepy and sleep state. The system is very efficient to detect the fatigue and control the train also train can be controlled if it cross any such signal by which the train may collide on another train.*

## 1. INTRODUCTION

In recent trends of development many safety technique and methods have been developed in detection of fatigue or drowsiness in the train for driver .The accidents are in increasing order of the train due to many reasons according to the survey and statistics 30% of the train accident are due to sleepy scenario detected for the train driver, many a time due to drowsy state train driver are not able to stop the train at the point when the train should of stop since driver is not able to take note of the signals. This may result in collision or train may go on the dead end track resulting in loss of many life and also involving damage to the expensive train equipments and property .Many incidents are the result of a driver failing to ensure that the train had stops at a stop signal due to falling asleep or might be died. Since from 19th century many methods have been developed to detect and prevent the fatigue or drowsiness in train drivers. These methods and technique involved warning and train stop systems. In India, a alarm system is used for providing the awaking mode. An buzzer or alarm sound is generated in driver's cabin if train passes and caution or stop signal given by control room. This system is called as (Automatic alarm system) AAS .The pervious approach of fatigue detection biological parameter for detecting fatigue can be shown in[1].

By adopting these features we are developing and integrated product by intelligently combine software and hardware used for detecting the current state for the train diver in this technique we are capturing the fatigue and motion as well as heartbeat rate providing alertness to the train driver. The processing of vision based detection of fatigue [2].

## 2. FATIGUE STATE DETECTION DURING DRIVING

The main objective of the system is that it is able to detect the fatigue or determine any status about the drivers position or state .In many scenarios driver may be suffering from heart attack or any other desperation caused due to heavy medication such position of the driver is also detected easily with the help of *heart rate detection system* this can be done by using heart beat sensor. Other than the heart beat sensor we are also using accelerometer to detect the motion of the face since the face motion of any normal has some range of movement motion but when an person is in the drowsy or fatigue mode the range of movement motion in the face change by this movement detection our system makes the technique more feasible to correctly detect the fatigue because previously developed many techniques had drawbacks of false fatigue detection. Different smart fatigue detection have been developed in area of fatigue[3].

There are many people who are travelling in train and every safety depends if driver is in proper state. Hence our developed system continuously keeps the track of the driver's state in the cabin and report it to the control room using the GSM module of the system

### 2.1 Visual Features:

The visual features are the most important factors used for detecting the fatigue .The head movement of driver and body movement are responsible since the person makes head movement in case of fatigue also no movement of driver states that driver is in sleepy mode or victim to any kind unconscious. The recent studied of visual features are based on the face recognition system[5]. The eye blinking is also an factor for detecting the fatigue of driver since the blinking rate of person awake differs from the blinking rate of person in sleepy state, as studied in[6].

### 2.2. Sleep /wake predictor model:

There are several process through which drowsy state of person can be detected for example "the Sleep/Wake Predictor Model (Akerstedt et al. 2004)",this model is called as Three- Process Model of Alertness, it is an model of computer that is based on timing of work or alertness of an person input taken by the model is work done by person in hours and sleep hours (amount of awake time and amount of sleep time) to detect the alertness of person and its performance based on the both specified parameter.

## 3. PROPOSED PLAN

The implemented system has build by using enhance technology hence the system contents different modules running simultaneously.

>There are total five modules in the proposed plan :

1. Module1:Image Capturing: This module capture the images and compare each image with the pervious image if images values are same then no fatigue has been detection if the image values differ from one another then fatigue is detected. We will be comparing the pre-captured images for testing the system.
2. Module 2:Face Movement Detection: This module detect the face movement and analysis the state of the train driver.
3. Module 3:Motion Detection: This modules is responsible to detect if driver is moving and performing some activity in cabin to determine the driver is in normal state

4. Module 4:Fatigue Detection: This module is able to detect the fatigue and avoid chances of the false fatigue detection in the system
5. Module 5 :Fatigue Alert: This module is responsible to develop an alert so that if driver is drowsy state he/she may come back to normal state by hearing the buzzer. This module basically implements the automatic alarm system[AAS]

The fatigue/inattention/drowsiness are very vague concepts. These terms refers a loss of alertness of vigilance while driving.

There have been many techniques and methods developed in the area of capturing visual human features ,most of these are related and based on facial recognition system to determine the face position ,movement of eyes and face, eye blinking frequency and many other. The eye blinking frequency and degree of open eyelid are the perfect indicators of sleepy state of a person[7]. In normal situation a person constantly keep on moving and blinking eye and there is an wide space between the eyelid when person is full wake.

In the state when person is sleepy the speed of blinking and opening frequency of the eyelid decreases. The moment and position of the driver's head angle in a normal situation, is a lifted up position and a similar moment relative to the driving is done. The fatigue state of the driver lead to frequent head movement this is movement of head is detected by the accelerometer

When person is in deep drowsy stage, the moment of head and nodding is extremely slow and the head moves itself completely relaxing position. The system developed is has enhanced features of correct fatigue detection in either case if the software fails to detect the fatigue or false fatigue the hardware is designed to be very efficient to detect the status of the driver and report is send by the help of GSM to the control room.

## 4. HARDWARE IMPLEMENTATION

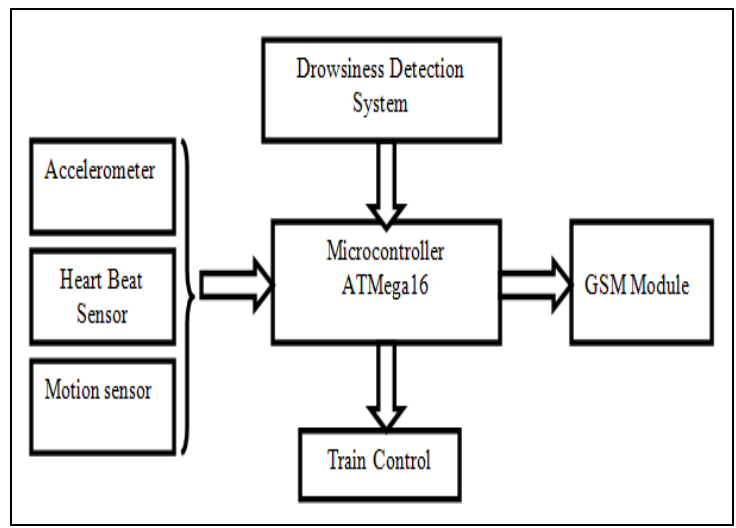

Figure : block diagram of proposed system

It is clear from the above block diagram that the webcam will detect the real time status of the driver for fatigue detection and generating alarm as per the condition. There will be three types of sensors will be attached with the system containing accelerometer, heartbeat sensor, Motion PIR Sensor, which will be used for real time status of the driver. If something gone wrong then microcontroller will directly send the status message of the driver their respective station master.

The developed system is made of combination of enhanced technologies like image processing, face detection, AVR studio ,Matlab .The software is responsible for detecting the motion and fatigue state while hardware is used for processing and transferring the signals from micro-controller.

### 4.1 Case1:

In general case of the proposed system is to detect the fatigue i.e. the drowsy state of the driver as shown in the figure accelerometer is activate to detect the continuous movement of the driver with the help of the motion sensor we are detecting the motion movement of the driver in case if the driver does not show any kind of movement the signals are immediately transfer to the micro-controller for further processing. When micro-controller receives the signal it immediately generate an alarm i.e. ringing sound to awake the driver. If the driver does not response then automatically the speed of the train is slowed to avoid further losses of signal if train is travelling on wrong track and the report regarding the status is transferred to control room through GSM.

### 4.2 Case 2:

In case 2 we describes the further part of hardware implementation , after the alarm is generated by the micro-controller and still the driver does not response to the system the heart-beat sensor will be activate to specify the death or alive state of the driver and the result are transferred to the control room authority through GSM. Second case used so that if software in rare case fails to detect the fatigue the hardware we had used i.e. the heart beat sensor will be activated for detecting the status of the driver. If driver does not response in particular time then case2 system is executed as this can be explain precisely in [4] about Time Nonintrusive Monitoring and Prediction of Driver Fatigue

The devices used to implement the system can be stated in brief as follows:

#### 4.2.3 Digital Subsystem

The digital system receives the signals from the different modules of the system and process accordingly .The digital system contain an micro-controller ATMega16 consisting of eight channels of 10bit A/D high accuracy.

## 5. SOFTWARE IMPLEMENTATION

Software used to develop the system are:

- Language use: Embedded 'C'
- Program writing and editing is done in AVR studio
- Micro-controller programming is done by using PROGISP
- Matlab

The proposed system has 2 types of software: one for image comparison with the help of Matlab using ATMega 16 micro-controller and another one for sending signals to control room detected by both fatigue detection system and heart beat sensor[8]. The execution of image processing is done in Matlab .The further processing and working of the Matlab can be studied from[9].
The real time function are perform in the following justified sequence.

1) Image capturing.
2) Image processing.
3) Processing of operation by ATMega 16.
4) Signals processing of heartbeat sensor .
5) The output of the system results in efficient fatigue detection and vehicle control system

The pulse measuring stage is very important for the HR calculation that is the main parameter .The signals processing is done by the ATMega 16 microprocessor and the response is forwarded to the train control room and GSM module .The response also constitute of the alarm system for the alertness of the driver.  There are different types of accelerometer motion sensing algorithm developed for motion detection[10].

The fatigue detecting and generation of alert system is an brief studied of the human behavior in driving the vehicle and automatic vehicle control system this is implement by using enhanced algorithms[11].

Improved Eye Tracking *Algorithm* to Be Used for Driver *Fatigue Detection System* Researcher have been successfully developed can give in areas[13].

# 6. CONCLUSION

The system has tried to overcome the drawbacks of previously developed fatigue detection system also tried to enhance the railways system by detecting the crossing of not notified signals due the abnormal state of driver in the cabin. Thus  we conclude that system if implemented on the application case will reduce the train accident about some extent.

The proposed system of fatigue detection in train driver leads to new enhancement in technology and in the area of computers since developed system is the combination of hardware and software .The system can be very effective in saving the life's of people in the train.

**Authors**

Miss. Monali Gulhane ,received the B.E GHRCEM Amravati, SGBA Amravati University ,Maharashtra, India and currently pursuing MTECH GHRCETW Nagpur, RTMN University Nagpur ,Maharashtra India. Her research area for post graduation includes image processing and artificial intelligence .

Prof.Prakash Mohod ,He is HOD in CSE Dept of GHRCETW Nagpur, Maharashtra ,India .He has received B.E (Computer Science & Technology), M.Tech (Computer Science & Engg).